\documentclass[twocolumn,floatfix,amssymb,amsmath,secnumarabic,nofootinbib, nobalancelastpage]{revtex4-1}    

\usepackage[pdftex]{graphicx} 
\usepackage{dcolumn}  
\usepackage{bm}       
\usepackage[usenames,dvipsnames]{color}
\definecolor{URLCOL}{rgb}{0,0.52,0.83} 
\definecolor{LINKCOL}{rgb}{0.05,0.5,0} 
\definecolor{CITECOL}{rgb}{0.25,0,0.48} 
\usepackage{epstopdf}
\usepackage[pdftex,bookmarks,breaklinks,bookmarksopen,bookmarksnumbered,colorlinks,
		    linkcolor=LINKCOL,linktocpage,citecolor=CITECOL,urlcolor=URLCOL,
			pdfpagemode=UseOutline,pdftex,pagebackref]{hyperref}
\usepackage{datetime}

\def\preprintlink{ \href{http://dft.uci.edu + PAPER REF}{title of paper} }
\def\preprinttext{Draft}

\usepackage{fancyhdr}
\makeatletter
\fancypagestyle{titlepage}
{
	\lhead{\textsc{\preprinttext\ from the \href{http://dft.uci.edu/publications.php}{Burke internal notes}}}
	\chead{}
	\rhead{\preprintlink}
	\lfoot{}
}
\makeatother
\def\preprintlink{ 
	\href{http://dft.uci.edu}
        {
dft.uci.edu}
	}

\usepackage{graphicx}
\usepackage{amsmath}
\usepackage{bm}
\usepackage{threeparttable}
\usepackage[normalem]{ulem}
\usepackage{array,multirow}
\usepackage{appendix}
\usepackage{epstopdf}
\usepackage{booktabs}
\usepackage{natbib}
\usepackage{feynmp}
\usepackage{threeparttable}
\usepackage{CJKutf8}
\definecolor{TITLECOL}{rgb}{0.1,0.2,0.7} 
\definecolor{PCOL}{rgb}{0.5,0.06,0.01} 
\definecolor{CHAPCOL}{rgb}{0,0.48,0} 
\definecolor{SECOL}{rgb}{0.1,0.2,0.7} 
\definecolor{CONTENTSCOL}{rgb}{0.1,0.2,0.7} 
\definecolor{SSECOL}{rgb}{0.25,0,0.48} 
\definecolor{SSSECOL}{rgb}{0.2,0.08,0.53} 
\definecolor{SHDCOL}{rgb}{0.4,0,0} 
\definecolor{ITMCOL}{rgb}{0.4,0,0} 
\definecolor{EXCOL}{rgb}{0,0.47,0.01} 
\definecolor{DEFCOL}{rgb}{0,0.42,0.01} 

\def\coloredtitle#1{\title{\textcolor{TITLECOL}{#1}}} 

\definecolor{URLCOL}{rgb}{0,0.17,0.43} 
\definecolor{LINKCOL}{rgb}{0.05,0.4,0} 
\definecolor{CITECOL}{rgb}{0.35,0,0.48} 

\definecolor{ngreen}{rgb}{0,0.48,0}

\def\sectable#1{
\addcontentsline{toc}{subsection}{~~Table: \textcolor{SSECOL}{#1}}
\begin{table}[h]
\caption{\bf \textcolor{SSECOL}{#1}}
}

\def\bea{\begin{eqnarray}}
\def\eea{\end{eqnarray}}
\def\ben{\begin{equation}}
\def\een{\end{equation}}
\def\benu{\begin{enumerate}}
\def\enu{\end{enumerate}}

\def\bei{\begin{itemize}}
\def\eei{\end{itemize}}
\def\beit{\begin{itemize}}
\def\eit{\end{itemize}}
\def\benu{\begin{enumerate}}
\def\enu{\end{enumerate}}

\def\n{n}

\def\sss{\scriptscriptstyle\rm}

\def\1var{(\bx_1...\bx\N)}

\def\half{\frac{1}{2}}

\def\bx{{x}}

\def\bj{{\bf j}}

\def\N{_{\sss N}}

\def\sph_int{ {\int d^3 r}}

\definecolor{SPECOL}{rgb}{0,0.47,0.01}
\definecolor{QUOCOL}{rgb}{0,0,0.2}
\definecolor{SHDCOLb}{rgb}{0.69,0.4,0.1}

\definecolor{SPEQ}{rgb}{0.01,0.4,0.05} %

\definecolor{SPEQv}{rgb}{0.45,0.05,0.45} %

\definecolor{SPEQb}{rgb}{0.01,0.1,0.65} %

\definecolor{SPEQr}{rgb}{0.57,0.05,0.1} %

\def\bay{\begin{array}}
\def\eay{\end{array}}
\def\bit{\begin{itemize}}
\def\beit{\begin{itemize}}
\def\eit{\end{itemize}}

\def\floor{\text{floor} }

\def\e{_{\rm e}}

\def\dd{~ \rotatebox{320}{\hspace{-5pt}\vbox to 5 pt {\hspace{-5pt} \hbox to 5pt {$\cdots$}}}\!\! }

\graphicspath{{Figures/}}
\begin{document}


\sf 
\coloredtitle{Explicit corrections to the gradient expansion for the kinetic energy in one dimension}
\author{\color{CITECOL} Kieron Burke}
\affiliation{Departments of Physics and Astronomy and of Chemistry, 
University of California, Irvine, CA 92697,  USA}
\date{\today, ArViv submission}
\begin{abstract}
A mathematical framework is constructed for the sum of the lowest $N$ eigenvalues of a
potential.  
Exactness is illustrated on several model systems (harmonic oscillator, 
particle in a box, and Poschl-Teller well).  
Its
order-by-order semiclassical
expansion reduces to the gradient expansion for slowly-varying densities, but 
also yields a correction when the system is finite and the spectrum discrete.
Some singularities can be avoided
when evaluating the correction to the leading term.
Explicit corrections to the gradient expansion
to the kinetic energy in one dimension are found which, in simple cases,
greatly improve accuracy.  We discuss the relevance to practical density functional
calculations.
\end{abstract}


\maketitle
\def\floor#1{{\lfloor}#1{\rfloor}}
\def\sm#1{{\langle}#1{\rangle}}
\def\dis{_{disc}}
\newcommand{\Z}{\mathbb{Z}}
\newcommand{\R}{\mathbb{R}}
\def\w{^{(0)}}
\def\w{^{\rm WKB}}
\def\II{^{\rm II}}
\def\hd#1{\noindent{\bf\textcolor{red} {#1:}}}
\def\hb#1{\noindent{\bf\textcolor{blue} {#1:}}}
\def\eps{\epsilon}
\def\ew{\epsilon\w}
\def\ej{\epsilon_j}
\def\upet{^{(\eta)}}
\def\ejeta{\ej\upet}
\def\tjeta{\tj\upet}
\def\bej{{\bar \epsilon}_j}
\def\ewj{\epsilon\w_j}
\def\tj{t_j}
\def\vj{v_j}
\def\F{_{\sss F}}
\def\xt{x_{\sss T}}
\def\sc{^{\rm sc}}
\def\al{\alpha}
\def\ae{\al_e}
\def\bj{\bar j}
\def\bz{\bar\zeta}
\def\eq#1{Eq.\, (\ref{#1})}
\def\cN{{\cal N}}

In the tens of thousands of density functional calculations published annually\cite{PGB16},
most 
employ the gradient of the density to estimate the exchange-correlation
energy of the Kohn-Sham equations\cite{KS65}.  Such approximations begin from the
gradient expansion of a slowly-varying electron gas\cite{Kc57}, which is then `generalized'
to an integral over an energy density with some function of the density
gradient\cite{LM83}.  The first such attempt came already in 1968 when Ma and Bruckner
showed that severe problems applying this gradient
expansion approximation (GEA) for the correlation energy to atoms could be
overcome by this procedure\cite{MB68}.
Since then, a variety of procedures and philosophies have been used to
construct such generalized gradient approximations (GGAs).  Some are
more accurate and popular in chemistry\cite{B88,LYP88}, while others work better for
(weakly correlated) materials\cite{PBE96,SRP15}.  This diversity reflects the ambiguity
in their derivation.   The older, simpler
local density approximation\cite{D30,KS65}, is unniquely determined by
the energy of the uniform electron gas\cite{CA80,PW92}.

Long ago, Lieb and Simon proved that, for any electronic system, the relative
error in Thomas-Fermi theory vanishes in a well-defined semiclassical limit
in which the particle number tends to infinity\cite{LS73,L76,LS77}.  Much work since
then studies corrections to this limit order-by-order, including extensions
of Thomas-Fermi theory\cite{E88}.  Such work is sometimes limited to atoms where spherical
symmetry simplifies the situation.  Englert beautifully summarized work with
Schwinger on this subject\cite{ES85,E88}.  However, this problem is complicated by
the interaction between electrons, the Coulomb attraction to nuclei, and
the complexities of semiclassics in three dimensions.

The present work studies the origin of the errors in applying the gradient
expansion in the simplest relevant case, namely the kinetic energy of non-interacting
electrons in one dimension.  This is not of quantitative relevance to
realistic electronic structure calculations.  The primary purpose
is the construction of a mathematical framework in which this question can
be directly addressed, and the errors of the gradient expansion explicitly 
identified and calculated in a systematic expansion in powers of $\hbar$.
We show that, for simple model cases, the formalism is exact, and also 
calculate the order-by-order expansion, finding great quantitative improvements
in energies when the corrections are accounted for.  We discuss the nature of
these corrections and how they might be incorporated in density functional
approximations.

Consider a symmetric potential $v(x)$, with zero chosen so that $v(0)=0$, and
which could tend to $D$, the well-depth, at large $x$.   Let
$\ej$ be the eigenvalues of the Schr\"odinger equation, using (Hartree) atomic
units (setting $m=\hbar=1$), and let
$M$ be the highest bound state if there is one.
The number staircase is
\ben
N_e(\eps)=\sum_{j=1}^M \Theta (\eps-\ej),
\label{Nmu}
\een
where $\Theta(x)$ is the Heaviside step function,
i.e., this is the number of states with $\ej < \eps$, and
$\Theta(0)=1/2$.  Next, consider a smooth
monotonic function $I(\eps)$ such that
\ben
I(\ej)=\bar j
\een
where $\bar j = j -1/2$.  The 1/2 comes from the
Maslov index for two turning points\cite{MF01}.  As $\hbar\to 0$,
a possible $I(\eps)$ is the classical action across the well, divided by $\pi$.
We also define $\eps(y)$ as the inverse of $I$, so that $\ej=\eps(\bar j)$.
Then
\ben
N_e(\eps)=\floor{I(\eps)+\half},
\een
where $\floor{x}$ is the highest positive integer less than $x$, so $N$ is the nearest
integer to $I$.
Next, define the periodic function
\ben
\sm{x}=x-\floor{x+\half},
\een
so that
\ben
N_e(\eps)=I(\eps)-\sm{I(\eps)}.
\een
To invert $N_e(\eps)$, we
turn on a temperature that is much smaller than any energy or
difference:
\ben
N_\beta (\eps)=\sum_{j=1}^M f(\beta (\eps-\ej)),
\label{Nbeta}
\een
where
$f^{-1}(x)=1+\exp(-x)$
and $\beta$ is inversely proportional to temperature.  We then define 
$\mu_\beta(\cN)$ as the inverse of $N_\beta$, $\cN \in  \R$.
For any finite
temperature, $\mu_\beta(\cN)$ exists and is well-defined.  We take
$\beta\to\infty$ at the end of the derivations and
stop mentioning the temperature explicitly.

\begin{figure}[htb]
\includegraphics[scale=.6]{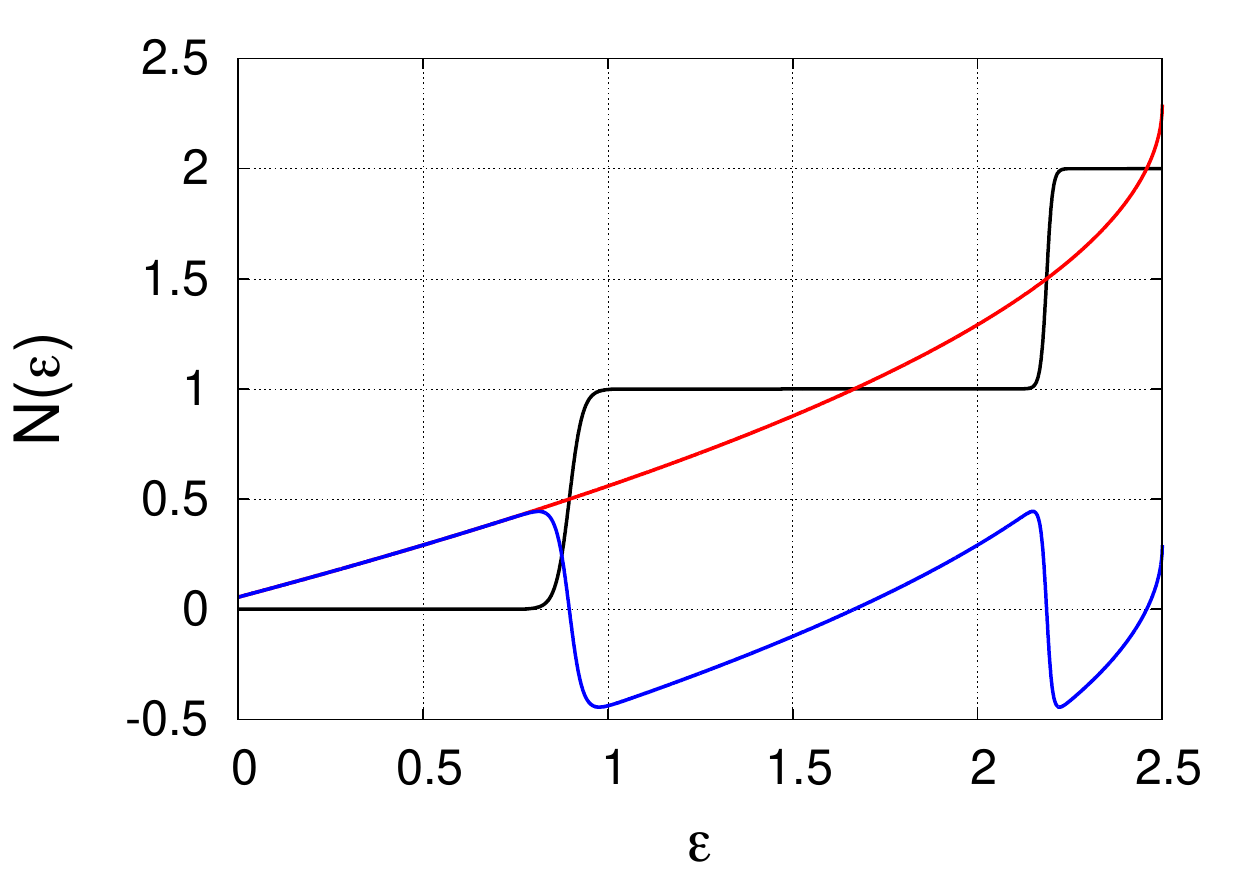}
\caption{Number staircase (black), smooth $I(\eps)$ (red), and their difference
$\sm{I(\eps)}$ (blue) for a Poschl-Teller well binding two states, rounded by 
a temperature of 0.01.}
\label{Neps}
\end{figure}
Fig. \ref{Neps} illustrates these functions for a Poschl-Teller well with
$D=5/2$ (see below).  The smooth $I(\eps)$ generates the staircase 
$N_e(\eps)$, whose steps are rounded by the temperature, making it
invertible.   The difference has a sawtooth shape, crossing zero at the eigenvalues,
so that $N_e=I$ when both are (half)-integers.

\begin{figure}[htb]
\includegraphics[scale=.6]{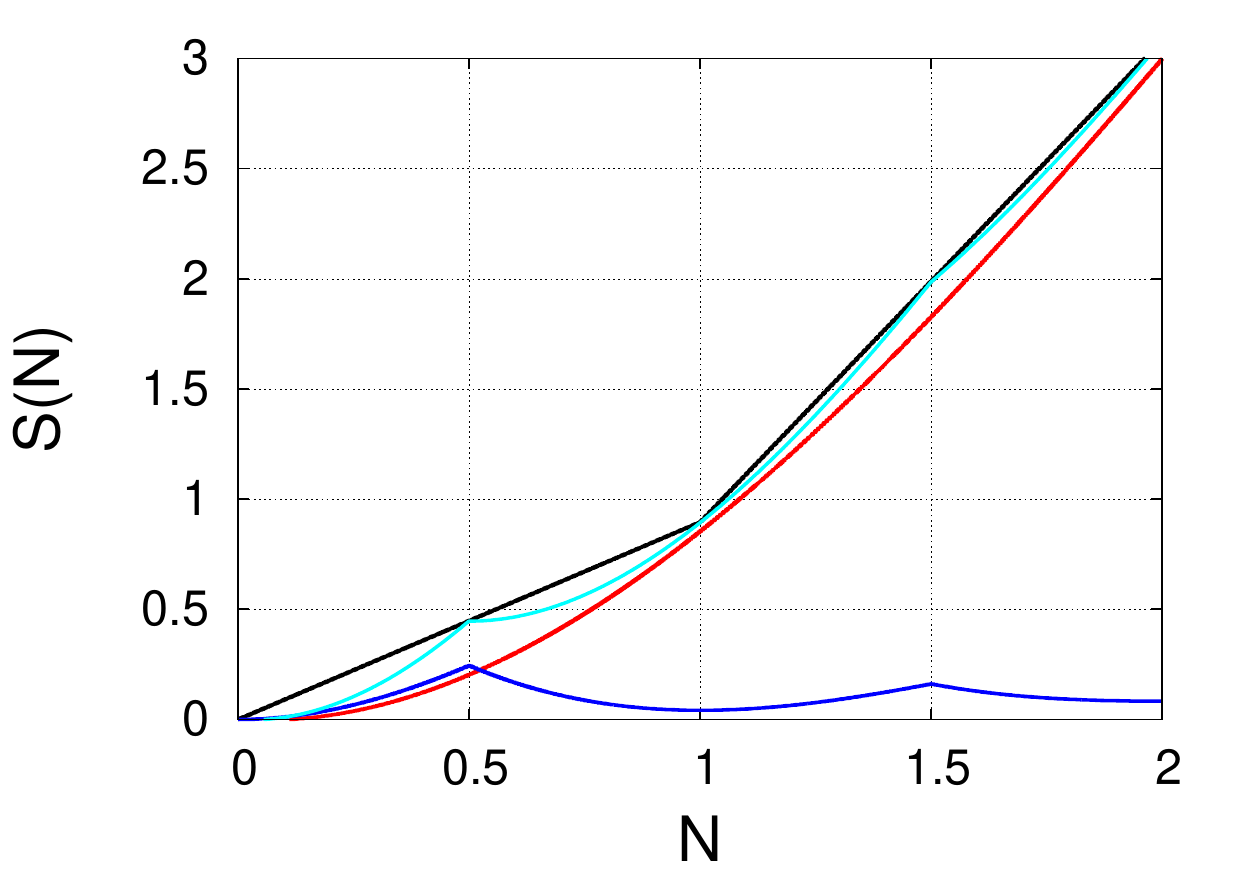}
\caption{Energy staircase as a function of $\cN$ (black), the integer interpolation of \eq{SintN} (cyan),
continuous contribution, $\cN \eps(\cN)-J(\cN)$ (red), and discontinuous contribution $G_{disc}$ (blue)
for the same
system as Fig. 1 (zero temperature).}
\label{SN}
\end{figure}
We wish to develop an expression for the sum of the eigenvalues, which would
be the total energy of $N$ same-spin fermions in the well.  Define
the energy staircase:
\ben
S(\eps)=\sum_{j=1}^M \ej\, \Theta (\eps-\ej),
\label{Seps}
\een
i.e., the sum of eigenvalues with energy below $\eps$.
It will be especially useful to consider:
\ben
G(\eps)=\int_{-\infty}^\eps d\eps'\, N\e(\eps') = \int_{-\infty}^\eps d\eps'\, \floor{I(\eps')+\half},
\een
and a well-known semiclassical result is\cite{BBb97}
\ben
S(\eps)= \eps\, N_e(\eps) - G(\eps),
\een
i.e., $N_e(\eps)$ uniquely determines the energy staircase.
But we really want $S$ as a function of particle number, $\cN$, which is
\ben
S(\cN)= S(\mu(\cN)) = \cN \mu(\cN) - G(\mu(\cN)).
\label{ScontN}
\een
This expression is well-defined for continuous values of non-negative $\cN$
(for non-zero temperature).  As the temperature goes to zero, it
becomes piece-wise linear, with changes of slope at integer values of $\cN$,
so that knowledge at integer values is sufficient to determine the whole function.
Both $\mu(\cN)$ and $G(\mu(\cN))$ have step-like features that combine to make this
happen, yielding
\ben
S(\cN)=S_N + (\cN-N)\, \eps_{N+1},~~~N=\floor{\cN}.
\een
Note that $\mu(\cN)=\eps(N)$
for integers, where $\eps=I^{-1}$, i.e., the discontinuous contributions in $\mu(\cN)$
vanish identically at integers, so they are not needed to find $S_N$.
In an obvious notation,
\ben
S(N)=N \eps(N) - G(N),~~~~N \in \Z.
\label{SNZ}
\een
Moreover, we change variables in the integration in $G$.  If $y=I(\eps)$,
then
\ben
G(N)=\int_0^N \frac{dy}{I'(y)}\, (y -\sm{y}),
\label{GN}
\een
where $I'=dI/d\eps$ and only the last term requires an integral over oscillations.
Eqs. (\ref{SNZ}) and (\ref{GN}) are a central result, providing the machinery to construct the sum of
the eigenvalues directly from $I(\eps)$, in continuous and discontinuous 
contributions.  
We define the first continuous term in $G$ as
\ben
J(\eps)=\int_{\eps(0)}^\eps d\eps'\, I(\eps'),~~~G\dis (N)=J(N)-G(N).
\een
The value of $\eps(0)$, negative in Fig. 1, is irrelevant to $G(N)$, as the step function
vanishes for arguments less than $1/2$, but not to $J(N)$ or $G\dis(N)$.
Because $J'=I$, $S_N$ is fully determined by $J(\eps)$.
Changing variables to $\eps(y)$ and integrating by
parts yields the more succinct\cite{BB20}
\ben
S\II(\cN) = \int_0^\cN dy\, \left\{ \eps(y) + \eps'(y) \sm{y} \right\},
\label{SintN}
\een
where II denotes integer-interpolation.
Fig. \ref{SN} plots quantities versus $\cN$ for a PT well,
showing that Eq. (\ref{SintN}) agrees with $S(\cN)$ only at (half)-integers.

A harmonic oscillator is instructive.  Here $I(\eps)=\eps/\omega$, so 
$\eps(y)=y\omega$, $I'=1/\omega$, $J=\omega N^2/2$, and $G\dis$ vanishes
because the average of $\sm{y}$ over one period vanishes if $I'$ is constant.
For a particle in an infinite well, 
\ben
I(\eps)=\frac{L{\sqrt{2\eps}}}{\pi}-\half.
\een
Then $\mu(N)=\pi^2(N+1/2)^2/(2L^2)$ and $1/I'=\pi^2(y+1/2)/L^2$.  Then
$J(\cN)$ is trivial to integrate but, because $I'$ varies, $G\dis$ does
not vanish:
\ben
G\dis (N)=\frac{\pi^2}{L^2}\int_0^N dy\, (y+\half)\, \sm{y}
\een
The constant term gives no contribution, while the integral over $y\sm{y}$ is
$-N/24$, producing the exact answer $S_N=\pi^2 N(N^2+3N/2+1/2)/(6L^2)$.
A less trivial example is provided by the Poschl-Teller well of depth $D$:
\ben
v(x) = D - D/\cosh^2(x).
\een
Writing $\ae={\sqrt{2D+1/4}}$, then 
\ben
I(\eps)=\ae - {\sqrt{2(D-\eps)}},
\een
yielding the eigenvalues
\ben
\ej = D -(\ae-\bj)^2/2,~~~~\bj < \ae.
\een
The simple result
$I' = 1/(\ae-I)$
makes the calculation easy, using the same integral over $\sm{y}$ as before,
giving
\ben
S^{PT}_N=\frac{\ae}{2}N^2 -\frac{N^3}{6}-\frac{N}{12}.
\een

So far, this result might be considered a simple tautology.  Its
real use comes when a semiclassical expansion is performed.  
We multiply $\hbar$ by a dimensionless number $\eta$, and consider the limit as
$\eta\to 0$.  Elementary analysis shows
\ben
\ejeta[v]=\eta^2\, \ej\left[\frac{v}{\eta^2}\right],
\een
and, as $\eta\to 0$,
\ben
I\upet[v](\eps) = I\left[\frac{v}{\eta^2}\right]\left(\frac{\eps}{\eta^2}\right)
= \frac{I^{(0)} [v]}{\eta} +\eta\, \Delta I^{(2)}[v]+...,
\een
where the expansion is known from WKB theory\cite{BO78}.
Here
\ben
I^{(0)}[v]=  \int_{-\infty}^\infty dx\ \frac{p(x)}{\pi}
\een
is the classical action divided by $\pi$,
$p(x)$ is the real part of the
local classical momentum, ${\sqrt{2(\eps-v(x))}}$, and
yields the (zero-order) WKB eigenvalues.
As $\eps^{(0)}(0)=0$,
\ben
J^{(0)} (\eps) = \int_{-\infty}^\infty dx\, \frac{p^3(x)}{3}.
\een

The semiclassical expansion for $S(\cN)$ {\em differs}
from the traditional WKB expansion.  The WKB expansion is an expansion of
individual eigenvalues in
powers of $\eta$, keeping $\eta\bj$ fixed, but the expansion of $S\upet(\cN)$
keeps $\eta \cN$ fixed.   One can both sum the WKB 
values to compare with $S_N$ and also compare $S_N-S_{N-1}$ with
the $N$-th WKB eigenvalue.
In general, these differ order-by-order (but 
infinite sums are identical).  
For example, summing WKB eigenvalues
for the PT well produces an additional $N/24$ relative to $S^{(0)}(N)$.
For $N > 1$, we expect the expansion of $S(N)$ to outperform the 
sum of WKB eigenvalues to the same order, as the semiclassical approximation
is used only at $N$, and not at each individual $j$ up to $N$, where it
should be less accurate.

The leading correction is trickier
to evaluate, due to a singularity as the turning points are
approached.  Define
\ben
B(\eps,a)=\int_0^{b(\eps)-a} dx\, \frac{v''}{p^3(\eps,x)}
\een
where $b(\eps)$ is the turning point at energy $\eps$, $0 < a < b(\eps)$.  
As $a\to 0$, a singularity develops which
must be cancelled:
\ben
\Delta I^{(2)}(\eps,a)=\frac{1}{12\pi} \left(B(\eps,a) + \frac{b''}{b'{\sqrt{2ab'}}}\right),
\label{dI2}
\een
and $\Delta I^{(2)}(\eps)$ is found by taking $a\to 0$.
This cumbersome procedure
can be elegantly avoided by an integral over a contour surrounding
the turning points\cite{BO78}. Higher-order terms involve even stronger singularities.
For a harmonic oscillator, Dunham\cite{D32} showed that all higher-order
terms are identically zero, so that WKB yields the exact answers.  Likewise for
a particle in a box,  as all derivatives of $v$ vanish, but
$\Delta I^{(2)}=1/(8{\sqrt{2D}})$ for PT.  

But we can instead evaluate the expansion for $J(\eps)$ and perform the
energy integration before the spatial integral\cite{MP56}.  
Consider the integral
\ben
A_\delta(\eps) = \int_\delta^\eps d\eps' 
\int_{-b(\eps'-\delta)}^{b(\eps'-\delta)} dx\, 
\frac{v''(x)}{p^3(\eps',x)},
\een
where $b(\eps)$ is the turning point at energy $\eps$.  For positive
$\delta$, this has no singularities, the order of integration
can be reversed, and the energy integral performed.  As $\delta\to 0$,
a singular term appears (of order $1/{\sqrt{\delta}}$) which cancels that
of \eq{dI2}.  Thus
\ben
J^{(2)}(\eps)=\lim_{\delta\to 0} \int_\delta^\eps\, d\eps'\,
I^{(2)}(\eps',v'(b)\delta)
=-\frac{1}{12\pi} \int_0^b dx\, \frac{v''}{p}.
\een

Because of the oscillation, $G\dis $ is already of higher-order
than the continuous terms.  Thus
\ben
G^{(2)}\dis(N)  = \int_0^N dy \sm{y}\, \frac{d\eps^{(0)}}{dy}
\een
Because of the periodicity of $\sm{y}$, only the endpoints contribute to the integral
as $\eta\to 0$, yielding
\ben
G^{(2)}\dis (N) = -\frac{1}{24} \, \frac{d \eps^{(0)}}{dy} \Big|_0^N.
\label{Goscv}
\een
Inserting all these pieces, and expanding $\eps(N)$ to second order in $J^{(0)}$
yields
\ben
\Delta S_N^{(2)} = - \Delta J^{(2)}(\eps^{(0)}(N)) + G^{(2)}\dis (N)
\een
which is $N^2/(16({\sqrt{2D}}))-N/12$ for the PT well.

Finally, we are ready to connect with density functional theory (DFT).
For 1d same-spin non-interacting fermions in a
slowly-varying potential in an extended system, there is a 
well-known expansion of both the density $n(x)$ and
kinetic energy $T$ in gradients of the potential\cite{SPb99}:
\ben
n^{(0)}(x)=\frac{p\F(x)}{\pi},~~~
\Delta n^{(2)}(x)=\frac{v''(x)}{12\pi\, p\F^3(x)},
\label{nGEA}
\een
and
\ben
T^{(0)}=\int dx\, \frac{p\F^3(x)}{6\pi},~~~~
\Delta T^{(2)}=\int dx\, \frac{v''(x)}{8\pi\, p\F(x)},
\label{TGEA}
\een
where $p\F(x)=p(\eps\F,x)$ and $\eps\F$ is determined by requiring the
density integrate to $N$.
Zero-order is Thomas-Fermi (TF) theory and 2nd order is the gradient expansion.
The combination $\cN\eps\F(\cN) - S(\cN)$ yields $T^{(0)}-2 \Delta T^{(2)}/3$,
agreeing with the semiclassical expansion of $J$, showing
\ben
S^{(2)}[v] = S^{GEA}[v] + G^{(2)}\dis [v],
\een
where GEA denotes (2nd-order) gradient expansion approximation.
The semiclassical expansion for $S_N$ reduces
to the gradient expansion for extended systems, where the sawtooth contribution
vanishes.   But the discontinuous contribution corrects GEA
to produce the exact leading-order correction to the local approximation for finite
systems.

\begin{table}[htb]
\begin{tabular}{|c|c|rrrr||c|rrr|}
\hline
\multicolumn{2}{|c}{}&\multicolumn{4}{|c||}{Error x 1000}&&\multicolumn{3}{|c|}{Error x 1000}\\
$N$&$S_N$&{TF}&{GEA}&2nd&4th&$\eps_N$&{WKB}&{$\Delta$TF}&{$\Delta$2nd}\\
\hline
1&  1.95&    69&  -42&    0.0& -0.000&1.95&111&69&0.0\\
2&  7.30&    110&  -83&    0.2& -0.001&5.35&82&41&0.1\\
3&   15.05&  122& -125&    0.4& -0.003&7.75&54&12&0.2\\
4&   24.20&  105& -166&    0.7& -0.005&9.15&25&-16&0.3\\
\hline
\end{tabular}
\caption{Energy sums for a PT well of $D=9.555$ and errors
of DFT approximations in milliHartree on left; errors in eigenvalues
on the right.  TF is the leading order, GEA is 2nd-order without
discontinuous contributions, 2nd includes them, and 4th-order does also.
WKB agrees with $S_N$ at 2nd-order.}

\label{T1}
\vskip -0.5cm
\end{table}
To see the performance of different approximations, some results for 
a generic Poschl-Teller well with $\ae=4.4$ are given in Table \ref{T1}.  The left side
gives errors for the sum of eigenvalues, the right for the individual levels.
The 2nd column of errors on the
left shows the result of the 2nd-order gradient expansion approximation (GEA),
which sometimes worsens results relative to TF theory.  But inclusion of the
correction reduces those errors by two orders of magnitude, yielding errors below one
milliHartree.  Addition of the next order reduces the errors to the microHartree range.
Deeper wells are even more favorable.
Switching to the right side, for this well
even the eigenvalues are better approximated
by differences in the sums within TF theory, but comparison
of their second-order contributions shows that,
in the asymptotic limit, WKB will have smaller errors than $\Delta$TF for the top
1/6 th of the levels.
\begin{table}[htb]
\begin{tabular}{|c|c|rrrr||c|rrr|}
\hline
\multicolumn{2}{|c}{}&\multicolumn{4}{|c||}{Error x 1000}&&\multicolumn{3}{|c|}{Error x 1000}\\
$N$&$S_N$&{TF}&{GEA}&2nd&4th&$\eps_N$&{WKB}&{$\Delta$TF}&{$\Delta$2nd}\\
\hline
    1& 0.50& 40&-40&  1& -0.08& 0.50& 82& 40&  1\\
    2& 1.50& -5&-78&  5& -0.32& 1.00& -4&-45&  4\\
\hline
\end{tabular}
\caption{Same as Table \ref{T1}, but with $D=1$.}
\label{T2}
\vskip -0.5cm
\end{table}
Table \ref{T2} repeats this calculation for $D=1$, which binds one particle
with energy $-1/2$ relative to the outside, and has a second level right at threshold.
This is extremely far from the semiclassical limit.  The trends are the same, but
errors in the full 2nd-order expansion are up to 5 milliHartree for the second level.
Here, WKB does better for the eigenvalue at the top of the well.

Lastly, consider density functionals.  Simply
invert \eq{nGEA} and insert the result into \eq{TGEA} to find\cite{SP99}
\ben
T^{GEA}[\n] = \frac{\pi^2}{6}\int dx\, \n^3(x) - \frac{1}{24}\int dx\, \frac{\n'(x)^2}{\n}.
\een
It is straightforward to convert \eq{Goscv} into a functional of the
TF density for the present circumstances, but care must be taken to
include the contribution at the lower endpoint.  For potentials with
a parabolic minimum at the origin:
\ben
G^{(2)}\dis [\n^{(0)}] = \frac{1}{24}\left(\frac{\pi^2}{C}-
\pi{\sqrt{\n\, |\n''|}}\Big|_{{\n^{(0)}(0)}}\right),
\label{Gdisn}
\een
where
\ben
C=\int_{-\infty}^\infty \frac{dx}{\n^{(0)}(x)}.
\een
Thus $G_{disc}^{(2)}$ contains both highly local and non-local contributions (integrals over local functionals),
and is to be added to the GEA.  Inserting the TF density for the PT well
correctly yields $N/24$.  Eq. (\ref{Gdisn}) looks like no local correction currently in the literature;
it has been derived, not devised.

The local density approximation applies to almost all situations.
The potential functional correction to GEA of \eq{Goscv} applies to many circumstances, such
as semi-infinite systems with surfaces, where the Maslov index differs, but 
must be generalized for e.g., multiple wells.  On the other hand, when converted
to a density functional, \eq{Gdisn}, the form of
the functional depends even further on the general class of
problem. 
For example the form differs from \eq{Gdisn} for $v(x)=|x|$.

A careful reader may note that no general prescription was given for finding $I(\eps)$.
For the DFT results, one needs only its well-defined asymptotic expansion.  For simple
model systems, the formulas used here suffice.  But 
adding any other function that vanishes at the eigenvalues generates 
equally viable candidates.  Different $I(\eps)$ yield different continuous
and discontinuous contributions, but still yield the exact sums.
The derivation using finite temperatures may appear cumbersome, but 
the formulas given generalize to thermal DFT\cite{M65,PY89}.

Many phenomena in DFT have a simple analog within this
1d world, as shown by two examples.
The first is the well-known inaccuracy of functional derivatives of 
reasonably accurate semilocal approximations for the energy\cite{KSB13}. 
This infamous misbehavior of the LDA XC potential leads to highly
inaccurate KS orbital eigenvalues.  The analog here is $N$-particle density
\ben
\rho(x) = \frac{\delta S_N[v]}{\delta v(x)}.
\label{rho}
\een
The archetype in 1d is the harmonic oscillator in TF theory, which yields
the exact eigenergies (and their sums), but whose density is highly inaccurate.
The local approximation is exact for the harmonic potential, but not
when small point-wise changes are made, as in \eq{rho}.
Only smooth changes in the potential should be expected to be correct in a local
theory (the first four moments (0-3) of the TF density of the oscillator are exact!).
Including the second-order correction yields densities that are singular
at the turning points.  This simply reflects the incompatibility of the order of
limits, by expanding in $\hbar$ before differentiating.

The second is the well-known difficulty of semilocal functionals when bonds are
stretched, a specific type of strong correlation\cite{CMY08}.  
Their failure
has been traced to a delocalization error, and related to curvatures of $E$ versus
$\cN$.  The same error shows up more strongly for the 1d kinetic energy.  For 
one particle in two well-separated identical potentials, half the density
ends up in each, leading to a factor of 4 reduction in the kinetic energy
relative to the one-well
result.
However, a model for the double well is
\ben
I_{double}(\eps) \approx I_{single}(\eps)+\sum_{j=1}^M \Theta_\beta(\eps-\ej)
\een
where $\beta$ is now a fixed large number, chosen to mimic the energy
splitting between even and odd levels.  The local approximation is
much smoother
\ben
I_{double}^{(0)}(\eps)=2\, I_{single}^{(0)}(\eps)
\een
and produces a huge overestimate for $j=1$.   In fact, the equivalence of
TF and WKB approximations breaks down, as there are now four turning points,
leading to ambiguities analogous to the symmetry dilemma\cite{PSB95} for stretched H$_2$.  This
occurs at the separation of the wells where the lowest
eigenvalue just touches the maximum in the potential.

The present work represents a culimination of a series of earlier
works\cite{ELCB08,CLEB11,RLCE15}
which focused on finding the density as a functional of the potential.
Its genesis was the failure of (very) improved uniform approximations
for the density to yield systematically improved kinetic energies\cite{RB17}.
The earlier results will prove useful when understood in the present context.
While model results are not directly relevant to realistic calculations, 
the understanding achieved from previous studies has already had
significant practical impact:  the derivation of the parameter in the B88 functional\cite{EB09},
an exact condition in PBEsol\cite{PRCV08} three exact conditions in the SCAN meta-GGA\cite{SRP15},
and the recently improved GGA correlation energy\cite{CCKBb18}.

This paper is aimed at the implications of this framework for DFT.
Work focused on the asymptotics of these sum formulas is in progress\cite{BB20}.
It is of tremendous
interest to apply this machinery in three dimensions.  The exact results
for sums apply to such a case, but the fluctuations around continuous
counterparts are far more complex.   Again, work on this subject is in progress.
Another area of great interest is their generalization to interacting systems.

I acknowledge funding from NSF (CHE 1856165), and the University of Bristol
for a Benjamin Meaker Professorship.  I
thank Attila
Cangi and Raphael Ribeiro, whose earlier work inspired this advance,
Michael Berry for instruction in semiclassics, and Chris Hughes for useful discussions.

\bibliographystyle{apsrev}
\bibliography{Master,BB}

\label{page:end}

\end{document}